# PROTOCELL SELF-REPRODUCTION IN A SPATIALLY EXTENDED METABOLISM-VESICLE SYSTEM

**Javier Macía[1] and Ricard V. Solé [1,2]**

[1] ICREA-Complex Systems Lab, Universitat Pompeu Fabra,

Dr Aiguader 80, 08003 Barcelona (Spain)

[2]Santa Fe Institute, 1399 Hyde Park Road, Santa Fe NM87501, USA

## Abstract

**Cellular life requires the presence of a set of biochemical mechanisms in order to maintain a predictable process of growth and division. Several attempts have been made towards the building of minimal protocells from a top-down approach, i.e. by using available biomolecules This type of synthetic approach has so far been unsuccesful, and the lack of appropriate models of the synthetic protocell cycle might be needed to guide future experiments. In this paper we present a simple biochemically and physically feasible model of cell replication involving a discrete semi-permeable vesicle with an internal minimal metabolism involving two reactive centers. It is shown that such a system can effectively undergo a whole cell replication cycle. The model can be used as a basic framework to model whole**

**protocell dynamics including more complex sets of reactions. The possible implementation of our design in future synthetic protocells is outlined.**

## INTRODUCTION

Membranes take part in many of the essential processes involved in the maintenance of cellular life (Albert et al., 2005; Lodish et al, 2005). They define the boundaries separating the inside world of chemical reactions and information from the outside environment. They play a very important role in exchanging substances with the external medium and in the growth and later cellular division. Current cells are very complex, as a result of a long evolutionary process, and have sophisticated mechanisms for cellular division regulation. However, the analysis of minimal cellular structures can contribute to a better understanding of possible prebiotic scenarios in which cellular life could have originated (Maynard Smith and Szathmáry, 2001) as well as in the design and synthesis of new artificial protocells (Rasmussen et al., 2004) .

A first approximation to a minimal cell structure is to consider a simplified model including the essential membrane physics defined in terms of the average behavior of a continuous, closed system involving some sort of simple, internal metabolism. Membrane division can take place spontaneously when membrane size goes beyond a critical value. In this case, the division process becomes energetically favored. However, cell division can be also actively induced through different mechanisms (Noguchi and Takasu, 2002). Understanding how these mechanisms trigger cell division can be very useful in designing synthetic protocells. Early work in this area was done by Rashevsky, who mathematically explored the conditions under which a single cell could experience a process of membrane expansion, deformation and splitting (Rashevsky, 1960). Given the lack of available computational resources at the time, the analysis was largely based on mathematical approximations.

An important component of a spatially-extended physically and chemically consistent model of protocell replication requires a minimal set of rules preserving the underlying chemistry and physics. In this context, Morgan et al. have developed well-defined (non-spatial) framework to the growth and division process as a non-stationary phenomenon (Morgan et al., 2004). In such approximation, the biochemistry is coupled to a spatially-implicit container, where the effects of cell geometry are introduced by means of scaling considerations. Some related approximations have been recently used in relation with the chemoton model (Munteanu and Solé, 2005). However, it would be also desirable to model cell replication under spatially-explicit conditions. This is particularly important in order to understand possible internal mechanisms triggering destabilization of closed membranes. Realistic models involving stochastic particle models (Ono and Ikegami, 1999) molecular dynamics (or dissipative particle dynamics) methods are not yet fully developed, and they are rather costly in computational terms (see Solé et al, 2006 and references therein). In these frameworks, each molecule (or some simplified representation of it as a particle) is explicitly modeled, with physical interactions taking place at the microscopic scale. This is especially relevant when dealing with nano-scale systems, but becomes less needed when dealing with lipid vesicles. Some current approaches to the synthesis of artificial cells are actually based on the use of a large vesicle as a container (Walde and Luisi, 2000; Luisi, 2000;

Oberholzer and Luisi, 2002; Hanczyc et al., 2003; Hanczyc and Szostak, 2004; see also Szostak et al., 2001 for a review and references therein).

Extensive experimental work has been developed using vesicles under a top-down approach (Luisi, 2002). This approach involves the use of molecules already present in living cells, including nucleic acids and complex enzymes. Such molecules would be enclosed within a liposome. The liposome would entrap the chosen molecules which could, under special conditions (typically unknown a priori) display cell-like properties. Different types of reactions, including the polymerase chain reaction (Oberholzer et al., 1995a) or the ribosomal synthesis of polypeptides (Oberholzer et al., 1995b) have been shown to occur inside these compartments. However, although such experiments indicate that complex chemical reactions can indeed occur within vesicles, no general framework exists on the potential conditions allowing such compartmentalized chemistry to trigger cell replication. An important drawback of these efforts has been the lack of a parallel development of simple theoretical and computational models able to capture the essential physical and chemical constraints consistent with cell self-reproduction. Such models would help driving the experimental design of minimal cells. Here we present a first step in this direction.

Among different possibilities of modeling spatial vesicles, the effects of time- and space-variable osmotic pressures seems one of the most suitable mechanisms inducing membrane division, since variable pressures can be generated by the internal metabolism, without additional external factors. What is required here is an active, non-equilibrium process able to trigger the growth the protocell. Provided that the interplay between metabolism and membrane geometric changes are able to trigger a symmetric deformation, cell division might be spontaneously achieved. In this paper we analyze how a simple metabolism can create such variable osmotic pressures and how the membrane becomes deformed under such differential pressures until completing division. For simplicity we restrict ourselves to a two-dimensional scenario.

## METABOLISM AND OSMOTIC PRESSURES

### Metabolic reactions:

Several possible implementations of a self-replicating cell can be constructed. Here we present one of them, leaving a general clarification to a future work. The object of study is a minimal cellular structure, formed by a closed continuous membrane enclosing a set of metabolic reactions. In our model, some of these reactions need the presence of two *metabolic centers* (enzymes) $E_1$ and $E_2$ adhered to the internal face of the membrane (see figure 1). This assumption is actually close to those performed by Rashevsky, although they can be relaxed (Macía and Solé, in preparation). These elements act as catalysts of these metabolic reactions:

(1.1)

$$\left.\begin{array}{l} R + E_1 \xrightarrow{k_1} 2G_1 + E_1 \\ R + E_2 \xrightarrow{k_2} 2G_1 + E_2 \end{array}\right\}$$

(1.2)

These enzymes might be specifically designed molecules including several catalytic centers (two identical molecules in this case).

The substance R is externally provided and available from the external medium. It is continuously pumped from a source located at the limits of the system, and can cross the membrane by diffusion, with permeability $h_R$. Similarly, the substance produced by metabolic reactions diffuses outwards by crossing the membrane with permeability $h_{G1}$.

The consumption and production of substances in the metabolic centers generates different flows, crossing the membrane in the normal direction. These flows depend on the difference of concentrations at each side of the membrane. Molecules tend to flow from regions of higher concentration to regions of lower concentration. Moreover, it is necessary to take into account the total water flow. The water flow depends on differential hydrostatic pressures inside and outside the membrane, as well as on the difference of solute concentrations: water tends to flow from regions of lower to regions of higher solute concentrations.

Since some of these reactions take place within a finite space domain (where the metabolic centers $E_1$ and $E_2$ are located) the distribution of the different substances is not uniform in the space. These spatially-localized reactions are the origin of non-uniform osmotic pressures along the membrane. On the other hand, due to these non-uniform osmotic pressures the membrane can become deformed and the location of the metabolic centers adhered to the internal side of the membrane can change along time. The combination of these two effects is the origin of time and space variable pressures. Furthermore, the membrane grows as a consequence of the continuous insertion of molecules or aggregates available from an external source, leading to changes in cellular volume and therefore to changes in concentrations.

All the concentrations are time and space-dependent. For a given molecule *j*, its concentration will be $c_j \equiv c_j(\mathbf{r},t)$, with $\mathbf{r}=(r_1,r_2)$ indicating the spatial coordinates. For notational simplicity, this dependence is not explicitly written. The concentration at each instant depends on the number of molecules $n_j$ and on the volume $V$. This dependence can be expressed as:

$$\frac{dc_j}{dt} = \frac{\partial c_j}{\partial n_j} \cdot \left. \frac{\partial n_j}{\partial t} \right|_V + \frac{\partial c_j}{\partial V} \cdot \left. \frac{\partial V}{\partial t} \right|_n \tag{2.1}$$

- Considering that $c_j(t)=n_j(t)/V(t)$, equation (2.1) can be written as:

$$\frac{dc_j}{dt} = \frac{1}{V} \cdot \left. \frac{\partial n_j}{\partial t} \right|_V - \frac{n_j}{V^2} \cdot \left. \frac{\partial V}{\partial t} \right|_n \tag{2.2}$$

- Finally the time evolution of the concentrations in the reaction-diffusion system coupled with the membrane is given by:

$$\frac{dc_j}{dt} = \left. \frac{\partial c_j}{\partial t} \right|_V - \frac{c_j}{V} \cdot \left. \frac{\partial V}{\partial t} \right|_n \tag{2.3}$$

The first term in the right hand accounts for the change in concentrations associated to changes in the number of moles $n_j$ inside the membrane, assuming constant volume. This term is described by the reaction-diffusion equations::

$$\left.\frac{\partial c_{G_1}}{\partial t}\right|_V = 2 \cdot k_1 \cdot c_R \cdot c_{E_1} + 2 \cdot k_2 \cdot c_R \cdot c_{E_2} + D_{G_1} \nabla^2 c_{G_1} \qquad (2.4)$$

$$\left.\frac{\partial c_R}{\partial t}\right|_V = R_0 - k_1 \cdot c_R \cdot c_{E_1} - k_2 \cdot c_R \cdot c_{E_2} + D_R \nabla^2 c_R \qquad (2.5)$$

Here $D_{G1}$ and $D_R$ are the diffusion coefficients of $G_1$ and R, respectively. In a first approximation, the model assumes that the values of these coefficients are the same inside and outside the membrane. $R_o$ is the constant supply rate of R in the external medium. Similarly, the second term in (2.3) account for the change in the concentration due to the volume changes, with a constant number of molecules $n_j$.

The flows crossing the membrane, are described by an additional set of equations. These equations account for the different interactions between all the elements of our system: water, solutes, and membrane (Kedem el al. 1958, Patlak et al. 1963, Curry 1984) . The following terms need to be considered:

- Water flow:

$$J_w = L_p\left(\Delta p - RT\left(\sum_{j=1}^{2}\sigma_j \Delta c_j\right)\right) \qquad (3.1)$$

where $L_p$ is the hydraulic conductivity of the water, $\Delta p$ is the hydrostatic pressure difference between the interior and the exterior, $R$ is the ideal gas constant, T is the temperature, and $\sigma_j$ is the solute reflection coefficient for the $j_{-th}$ substance (0 for a freely permeable solute, and 1 for a completely impermeable solute). Here $\Delta c_j = c^e_j - c^i_j$ is the concentration difference of the $j_{-th}$ substance at both membrane sides (exterior minus interior). The index $j$ corresponds to the different substances: $j = G_1$, R.

- Solute flows: For dilute solutions, the solute-solute interaction can be no considered. For each different substance the flow is given by:

$$J_j = h_j \Delta c_j \left(\frac{P_e^j}{e^{P_e^j} - 1}\right) + J_w (1 - \sigma_j) c_j^i \qquad (3.2)$$

$h_j$ is the permeability of the substance $j$ (defined as the rate at which molecules cross the membrane), and $P^j_e$ is the so called Peclet number, given by:

$$P_e^j = \frac{J_w (1 - \sigma_j)}{h_j} \qquad (3.3)$$

The right hand of equation (3.2) is the sum of convective and diffusive components of solute flow. The first term is the diffusive flow, which depends on the difference of concentrations. The second term if the convective flow, and accounts for the amount of solute carried across the membrane by net water flow:.

Cell volume changes due to both net water flow as well as to the growth of the membrane:

$$\frac{dV}{dt} = J_w A \tag{4.1}$$

If the composition of the external solution does not change over time, the rate at which the externally provided compounds are incorporated into the membrane can be considered proportional to its area:

$$\frac{dA}{dt} = \frac{\ln 2}{T_d} A \tag{4.2}$$

where $T_d$ is the time taken for the membrane to double its area.

**Osmotic pressures, surface tension and bending elasticity:**

The set of equations (2.1-4.2) describes the dynamics of the system. To explicitly model the membrane shape evolution, it is necessary to consider the effect of the different flows, in terms of local pressure, at each point of the membrane. The flows of the different substances generate different osmotic pressures. At each point the osmotic pressure value $P^o_j$ generated by the substance $j$ depends on the different concentrations of this substance at each side of the membrane:

$$P_j^o(r,t) = k_j \cdot (c_j^i(r,t) - c_j^e(r,t)) \tag{5}$$

where $k_j$ is constant. For the particular case of very low concentrations, we have $k_x \approx R \cdot T$, where R is the ideal gas constant and T is the temperature (if the concentrations are expressed in mols/liter). For many solutes, the osmotic pressure is not proportional to their concentrations. In these cases the osmotic pressure is empirically fitted to a polynomial function of the concentration. Thus, equation (5) is an approximation for small solutes.

The osmotic pressure at one point **r** of the membrane at time t can be calculated by adding the pressure generated by each substance, as follows:

$$P_t^o(r,t) = \sum_j k_j \cdot (c_j^i(r,t) - c_j^e(r,t)) \tag{6}$$

Finally it is necessary take into account the contribution of the surface tension and the bending elasticity to the total pressure. This contribution is described by the following expression:

$$P_{total}(r,t) = P_t^o(r,t) + \frac{2\gamma}{R_S(r)} + \frac{\kappa}{[R_S(r)]^2}\left(\frac{1}{R_S(r)} - \frac{1}{r_o}\right) \tag{7}$$

where $\gamma$ is the surface tension coefficient, and $\kappa$ is the elastic bending coefficient. Equation (7) is valid only for a 2D model. However, a more detailed model, taking into account the chemical composition of the membrane, needs a more precise estimation for $\gamma$ (White 1980). Here $R_s(r)$ is the local radius of curvature. This value is given by the radius of the circumference with the best fit to the real membrane curvature in a local environment of the point r. Finally $r_o$ is the spontaneous radius of curvature.

## SIMULATION MODEL

In this section we present and analyze the reaction-diffusion system coupled with membrane shape evolution. There are different methods to study the evolution of the membrane shape under different conditions, such as the so called Phase Field Method (Du et al. 2005 and references therein). In general these methods are based on the assumption of global constrains, i.e. the minimization of the elastic bending energy. One of the goals of this paper is to introduce a method to track membrane surface without any assumption a priori on of global conditions. Our method only takes into account the local effects of the different pressures acting at each point. As will be shown below, by considering these local effects, the expected minimization of the elastic bending energy emerges. On the other hand, a second goal is to analyze if a non-uniform pressure distribution along the membrane is enough for a controlled membrane deformation to occur, until complete division takes place. The rules defining our model are presented below.

### The reaction-diffusion system

The reaction-diffusion system described by equations (1.1-1.2) can be calculated by using a discrete approximation using both discrete time and space (Schaff et al., 2001, Wiemar et al., 1994, Patankar, 1980). With such model it is possible to properly model the membrane behavior. Figure 2a indicates how this discrete approximation can be performed. The available space is divided into discrete elements of area $dS=dx \cdot dy$. Each discrete element is identified by its column and row (i,j). There are tree types of elements:

- Discrete internal elements, which cover all the area inside the membrane.

- Discrete external elements, which cover all the external area.

- Discrete membrane elements, which cover the membrane.

To construct the discrete approximation to the real membrane, the membrane elements must be in contact with both internal and external elements, and all the elements must define a closed system. Not all the elements in contact with the membrane can be employed in this membrane approximation since they can be in contact only with one of the internal or external elements, but not to both.

To perform the calculations, it is useful to separate the concentrations: for a discrete internal element located at the site (i,j) at time *t*, the concentration of the $j_{\text{-th}}$ substance is indicated as $\hat{c}^{i}_{j}(i,j,t)$ and for an external element is indicated as $\hat{c}^{e}_{j}(i,j,t)$. These are the concentration values used in the finite differences approximation for equations (2.1-2.5). Depending on the discrete elements, the concentrations values must be:

- Discrete internal elements: $\hat{c}^e_x=0$ y $\hat{c}^i_x\neq0$

- Discrete external elements: $\hat{c}^e_x\neq0$ y $\hat{c}^i_x=0$

- Discrete membrane elements: $\hat{c}^e_x\neq0$ y $\hat{c}^i_x\neq0$

**Membrane shape characterization**

This model assumes that the membrane is a closed and continuous boundary. To track the membrane shape evolution coupled with the metabolism described by equations (1.1-1.2) it is necessary to define a set of *characteristic points* $Q_k$ along the membrane. The shape of the membrane at each time is determined by the spatial position of these points. The shape of the membrane between two neighboring characteristic points can be obtained from a linear interpolation. To make a correct choice of these characteristic points it is necessary to take one point in each discrete membrane element (figure 2). Initially the characteristic point for each discrete membrane element is chosen in the middle of the segment that crosses the discrete membrane element.

**Flow terms**

Using our discrete approximation to the membrane structure, instead to a ideal continuous membrane, we need to introduce some corrections in the calculation of discrete flows. First, we need to determine the normal direction to the membrane at each discrete membrane element. The normal direction associated to the membrane element (i,j) is defined by the angle $\Phi_{(i,j)}$ between the normal to the ideal membrane at the characteristic point and the horizontal axis (see figure 2a.).

When the calculations are performed on a discrete lattice, due to the diffusion process each element exchanges molecules with just its nearest neighbors, as figure 2b shows. Let us indicate as $g_{(i,j)-(l,m)}$ the exchanged flow between elements (i,j) and (l,m) if both are internal or external elements. If the element (i,j) belongs to a discrete membrane element, the exchanged flow is $g_{(i,j)-(l,m)} \cdot A_{(i,j)-(l,m)}$ with (Schaff et al., 2001):

$$A_{(i,j)-(l,m)} = \begin{cases} \cos\Phi_{(i,j)} & if \quad i = l \\ \sin\Phi_{(i,j)} & if \quad j = m \end{cases} \qquad (8)$$

In this situation, the estimated flows across the membrane need to be corrected using (8).

**Membrane Deformation**

In the discrete approximation, at each characteristic point $Q_k$ the difference of concentrations between both sides is the difference of concentrations $\hat{c}^i_j$-$\hat{c}^e_j$, for the substance *j*, associated to the discrete membrane element containing the characteristic point. This difference of concentrations at both sides creates an osmotic pressure. Under certain conditions these pressures are enough to deform the membrane. To simulate this effect it is necessary to assume that each characteristic point can change its position under influence of these pressures. In a first approximation, the displacement of each point is proportional to the total pressure described by (7).

$$Q_k(t+\Delta t) = Q_k(t) + b \cdot P_{total}(t, Q_k) \tag{9}$$

with *b* being a constant and Δt the discrete time interval used in the computation. The value of *b* cannot be arbitrary (see the results section for more details). With the new locations of $Q_k$ it is possible to generate the new shape of the membrane using a linear interpolation. Once this process is completed, it is necessary to define which discrete elements are internal, external or membrane elements again.

The membrane growth is described by equation (4.2). This growth must be enough to guarantee that the membrane remains closed all the time. The membrane size defined by the characteristic points locations must agree with the size predicted by (4.2).

**RESULTS**

Taking into account the previous local effects, the global behaviour of the membrane should correctly described. To test the validity of the results provided by the rules of the model, different analysis have been performed. In the following sub-sections, we study three key features of out protocell model, namely:

(a) the time changes of membrane shape

(b) the volume growth under non equilibrium conditions

(c) the process of cell division.

These three aspects of the model capture the essential features of the underlying physics and chemistry and allow testing the realiability of the model predictions.

**Free membrane evolution.**

A first simple test of the correctness of our approach is given by the analysis of the membrane relaxation dynamics. Some methods for membrane shape evolution, such as the Phase Field Method, are based in the physical principle of energy minimization. The surface must evolve freely towards shape configurations of minimal elastic bending energy. This energy is given by (Du et al.2005):

$$E = \int_{\Gamma} \left[ a_1 + a_2 (H + c_0)^2 + a_3 G \right] dS \qquad (10)$$

where $a_1$ is the surface tension, $a_2$ is the bending rigidity, $a_3$ is the stretching rigidity, and $c_0$ is the spontaneous curvature. H is the mean curvature and G is the Gaussian curvature. The integral is over the entire membrane surface Γ.

In our method global conditions on energy minimization are not imposed *a priori*. However, to be physically coherent, this global behaviour must emerge from the local rules describing the model. In particular, we should expect to observe a predictable shape change, starting from an arbitrary initial state, converging towards a circular configuration. Figure 3 shows the evolution of the elastic bending energy through time. Starting from an arbitrary shape, the membrane shape relaxes towards configurations of minimal energy.

**Volume growth and coefficient of displacement estimation.**

Consider a membrane with spherical symmetry in osmotic equilibrium with its surrounding medium. If the external concentration decays rapidly, this produces a difference in the osmotic pressures between the internal and external medium, and the water flows inwards (the membrane is impermeable to the solutes). This water flow produces an increase of the membrane volume given by:

$$\frac{1}{A}\frac{dV}{dt} = L_p \Delta p \tag{11}$$

where A is the membrane area, $L_p$ is the hydraulic conductivity and $\Delta p$ is the osmotic pressure difference. As the cell volume increases as a consequence of the water flow, the membrane area increases too. Let us assume that the increase in area is only due the elastic expansion (i.e. no new molecules or aggregates are incorporated to the membrane). Using the spherical symmetry, from (11) the increase in the membrane area can be approximated by (Wolfe et al. 1986):

$$A(t) \approx A_F - (A_F - A_I) \cdot e^{-\frac{t}{\tau}} \tag{12}$$

with $A_F$ the final area, $A_I$ the initial area, and $\tau=(A_F-A_I)/8\pi L_p R_I$. Here, $R_I$ is the initial radius value.

In order to test our method, the simulation starts with a spherical membrane, with a radius of 8 μm, in equilibrium with its surrounding medium. The internal concentration is 0.1 mol/l. When external concentration decreases rapidly to 0.07 mol/l the water flows inwards the membrane to equilibrate the chemical potentials (values of hydraulic conductivity, surface tension and elastic coefficient from table I). During all the process the spherical symmetry remains. Figure 4 show the results obtained by numerical simulation using our method, compared with the analytic result obtained by equation (12). This simulations have been performed with different values of the displacement coefficient *b* (see equation 9). The best fit is obtained with $b$=$1.57 \cdot 10^{-13}$ $cm \cdot Pa^{-1}$ with the parameters employed.

These results suggest that the local rules defining our method capture the essential physics of membrane deformations.

Our simulations show that, under certain conditions, osmotic pressures can deform and in some cases eventually split the membrane. The osmotic pressure values of a given substance depend on the different concentrations at each side of the membrane. These different concentrations basically depend on two parameters: the permeability and the different values of the diffusion coefficient inside and outside the membrane. In these simulations the diffusion coefficient is the same in both sides, therefore the permeability becomes the fundamental parameter.

Table 1 shows the parameter values used in the simulations. These have been performed on a 128x128 discrete, square lattice. Initially, the metabolic centers $E_1$ and $E_2$ are attached to the membrane in two opposite locations. $G_1$ is produced in both metabolic centers and diffuses through the membrane pushing it outwards. Moreover, R comes form the external medium and pushes the membrane inwards.

Figure 5 shows the pressure distributions along the membrane at different time steps of the evolution of the membrane-metabolism system. The pressures associated to $G_1$ favors membrane expansion where they are higher. This occurs where $E_1$ and $E_2$ are located. Figure 6 shows the concentration profiles of $G_1$ inside the membrane. When the narrowing is enough, two independent membranes enter into contact. This qualitative change depends on the distance between the characteristic points in the narrowed zone (see Appendix I).

During the expansion-narrowing process, the membrane size necessary to ensure that the cell boundary is continuous and closed increases. Such increase is due the incorporation of externally provided molecules, following equation (4.2). Figure 7 shows the good fitting between the required size needed to close the membrane along the characteristic points and the real size of the membrane, as calculated from (4.2).

## DISCUSSION

One of the greatest challenges of synthetic biology is the construction of simple protocells able to self-reproduce themselves. Efforts in this direction have used liposomes as the containers and special reactions that have been shown to occur inside such vesicles. Although the feasibility of such reactions is a very positive result, these bioreactors have failed so far to display self-reproduction. Only in one case (Oberholzer et al., 1995c) molecular self-replication was shown to be coupled to the whole vesicle reproduction. A systems approach might help understanding the potential scenarios allowing minimal synthetic cells to undergo a cell cycle. This requires both a simple, physico-chemically feasible design as well as an appropriate computational implementation.

We have presented a minimal cellular model formed by a continuous closed membrane with a simple, enzyme-driven internal metabolism. It has been shown that the effect of variable osmotic pressures, under certain conditions, can be a regulatory mechanism for the division process. These osmotic pressures are generated by the internal cellular metabolism, consistently with some old theoretical predictions (Rashevsky, 1960). The behavior of the membrane under the active metabolism is driven by variable osmotic pressures which can be very relevant to the synthesis of artificial protocells, as well as in understanding some prebiotic scenarios, where the sophisticated division mechanisms of the current cells were not present. In this context, the set of reactions defining the internal metabolism can be arbitrarily generalized, thus opening the door for many different types of membrane-metabolism couplings.

The metabolism analyzed need two metabolic centers $E_1$ and $E_2$ linked to the internal side of the membrane. This can be achieved using transmembrane proteins with the appropriate reaction centers able to catalyze several reactions simultaneously (two in our example). After cellular division, each daughter cell has only one metabolic center. At this point, the division process cannot start again without the replication of the metabolic center. This is a limitation as far as it would be desirable that protocells keep reproducing indefinitely. However, our simulations suggest that a metabolism where the non-uniform concentration distribution arise form a spatiotemporal pattern (as in Turing patterns) without specifically located metabolic centers could be an efficient self-replication mechanism for minimal cells. Moreover, our results and design are compatible with available understanding on molecular reactions and vesicle dynamics. The parameters used in our implementation as well as the predicted behavior are consistent with physically and chemically reasonable limits, providing support for our design as a feasible model of minimal protocell.

## ACKNOWLEDGMENTS

The authors thank Dr. Carlos Rodríguez-Caso and the rest of the members of the Complex System Lab for useful discussions. This work has been supported by EU

PACE grant within the 6th Framework Program under contract FP6-0022035 (Programmable Artificial Cell Evolution), by McyT grant FIS2004-05422 and by the Santa Fe Institute.

## APPENDIX I: Membrane division

A final rule is required to effectively generate a separation between two daughter cells. Since the membrane is actually a continuous medium, this model could not allow the total break of the membrane into two independent closed membranes unless some additional change is introduced. The membrane split into two independent membranes is a singularity. This singularity can be easily introduced in the simulations (Appendix I).

After a deformation process, the new membrane shape is determined by interpolating between the different characteristic points of the membrane. This interpolation is performed in a clockwise direction. Given one characteristic point $Q_k$ all the others points $Q_m$ have a set of associated weight values. These weight values are calculated by using a distribution:

$$W(d_{Q_mQ_k}) = \frac{-A}{(d_{Q_mQ_k})^{12}} + \frac{B}{(d_{Q_mQ_k})^{6}} \qquad (10)$$

where $d_{QkQm}$ is the distance between both points. This function has a behavior similar to Lennard-Jones potential, frequently employed to describe the boundary between lipids in a membrane. Figure A.1 shows the weight profile used here. Given one characteristic point $Q_k$ the interpolation will be performed between this point and the point $Q_m$ with higher value of $W(d_{QkQm})$ in the clockwise direction. Figures A.2a and A.2b show how this process can take place. In the figure 4a, with a spheroid shape, the interpolation process starts at the characteristic point labeled $Q_i$ (this initial point is arbitrary). The next point with higher weight value, in the clockwise direction, is $Q_{i+1}$, so the interpolation between $Q_i$ and $Q_{i+1}$ will be performed. The next interpolation will be between $Q_{i+1}$ and $Q_{i+2}$, and so on. Figure A.2b shows a membrane with a very high deformation. In the narrow neck zone the interpolation between $Q_i$ and $Q_{i+1}$ will be performed, but from $Q_{i+1}$ the interpolation is possible with the point labeled $Q_{i+2}$ or with the point labeled $Q'_{i+2}$, depending if $W(d_{Qi+1,Qi+2})>W(d_{Qi+1,Q'i+2})$ or $W(d_{Qi+1,Qi+2})<W(d_{Qi+1,Q'i+2})$. If the interpolation is between $Q_{i+1}$ and $Q_{i+2}$ the simulation works with one deformed membrane, but if the interpolation is between $Q_{i+1}$ and $Q'_{i+2}$ the simulation assumes two closed membranes, one in the top half and other in the bottom half, in contact.

This is a *qualitative* rule in order to impose membrane splitting. Without this rule the membrane behaviour is the same but the final splitting cannot take place due the continuous nature of the membrane assumed in this model.

## References

Alberts, B. et al. 2005. *Molecular Biology of the Cell*. 4th edition. Garland, New York

Bozic, B., Svetina, S.. 2004. A relationship between membrane properties forms the basis of a selectivity mechanism for vesicle self-reproduction. Eur. Biophys. J. **33**: 565-571

Curry FE. 1984. Mechanics and Thermodynamics of Transcapillary Exchange. In: *Handbook of Physiology, Section 2, The Cardiovascular System,* edited by Renkin EM and Michel CC. Bethesda: American Physiological Society, p. 309-374.

Du, Q., Liu, C., Wang, X. 2005. Simulating the deformation of vesicle membranes under elastic bending energy in three dimensions. *Journal of Computational Physics* 212, 757-777.

Hanczyc, M. M., and Szostak, J. W. 2003. Experimental Models of Primitive Cellular Compartments: Encapsulation, Growth, and Division. *Science* 302, 618 - 622

Hanczyc, M. M. and Szostak, J. W. 2004. Replicating vesicles as models of primitive cell growth and division. Curr. Opin. Chem. Biol. 8, 660-664.

Kedem, O., and Katchalsky, A. 1958. Thermodynamic analysis of the permeability of biological membranes to non-electrolytes. *Biochim. Biophys. Acta* **27,** 229–246.

Lodish, H. et al., 2005. *Molecular Cell Biology*. 4th edition. Freeman, New York.

Luisi, P. L. 2002. Toward the engineering of minimal living cells. Anat. Record 268, 208-214.

Maynard Smith, J., Szathmáry, E. 2001. *The Major Transitions in Evolution*. Oxford University Press.

Morgan, J. J., Surovtsev, I. V., Lindahl, P. A. 2004. A framework for whole-cell mathematical modeling. J. Theor. Biol. **231**, 581-96.

Munteanu, A. and Solé, R. V. 2006. Phenotypic Diversity and Chaos in Protocell Dynamics. J. Theor. Biol. In press.

Noguchi, H., Takasu, M., 2002. Adhesion of nanoparticles to vesicles: a Brownian dynamics simulation. Biophysical Journal. 83, 299-308

Oberholzer, T., Albrizio, M. and Luisi, P. L.2005a. Polymerase chain reaction in liposomes. Chem Biol. 2, 677-682.

Oberholzer, T., Nierhaus, K. H. and Luisi, P. L. 2005b. Protein Expression in Liposomes. Biochem. Biophys. Res. Comm. 261, 238-241.

Oberholzer, T., Wick, R., Luisi P.L. and Biebricher, C.K. 1995c. Enzymatic RNA replication in self-reproducing vesicles: an approach to a minimal cell. 207, 250-257.

Oberholzer, T., Luisi, P.L. 2002. The use of liposomes for constructing cell models. J. Biol. Phys. 28, 733-744

Ono, N. and Ikegami, T. 1999. Model of Self-Replicating Cell Capable of Self-Maintenance. In: Lecture Notes In Computer Science; Vol. 1674, 399-406. Springer, London UK.

Patlak, C. S., Goldstein, D. A. and Hoffman, J. F. 1963. The flow of solute and solvent across a two-membrane system. Journal of Theoretical Biology 5, 426-442.

Patankar, S., 1980. *Numerical Heat Transfer and Fluid Flow*. Taylor and Francis, Washington D.C.

Rashevsky, N. 1960. *Mathematical Biophysics. Physico-matemathical foundations of biology*. Vol I. Dover Publications, Inc. New York, USA.

Rasmussen, S., Chen, L., Deamer, D., Krakauer, D. C., Packard, N. H., Stadler, P. F. and Bedau, M. A. 2004.Transitions from Nonliving to Living Matter. **303**, 963 - 965

Schaff, J. C., Slepchenki, B. M., Yung-Sze Choi, Wagner, J., Resasco, D., Loew, L. M., 2001. Analisis of nonlinear dynamics on arbitrary geometries with the Virtual Cell. Chaos **11**, 115-131.

Solé, R. V., Macía, J., Fellermann, H., Munteanu, A., Sardanyés, J. and Valverde, S. 2006. Models of protocell replication. In: *Protocells: bridging living and non-living matter*. Steen Rasmussen et al., editors. MIT Press.

Szostack W., Bartel, D. P. and Luisi, P. L. 2001. Synthesizing life. Nature **409**, 387-390

Walde, P. and Luisi, P. L. (eds) 2000. *Giant vesicles*. John Wiley, Canada.

White, S. H., 1980. Small phospholipids vesicles: Internal pressure, surface tension, and surface free energy. Proc. Natl. Acad. Sci. USA. **77**, 4048-4050.

Wiemar, J.R., Boon, J.P., 1994. Class of cellular automata for reaction-diffusion systems. Phys. Rev. E **49**, 1749-

Wolfe J., Dowgert, M.F., Steponkus, P.L. 1986. Mechanical study of the deformation and rupture of the plasma membranes of protoplasts during osmotic expansions. J. Membrane Biol. 93, 63-74.

## FIGURE CAPTIONS

**Figure 1.** Schematic representation of the minimal cell. The structure is formed by a closed continuous membrane. Adhered to the internal side of the membrane there are two enzymes $E_1$ and $E_2$. These enzymes are the catalyst of a part of the metabolic reactions. When a molecule of R is in contact with $E_1$ this molecule breaks into a pair of molecules $G_1$, $S_1$. The same occurs when R is in contact with $E_2$. In this case R is broken into $G_2$ and $S_2$. $G_1$ and $G_2$ produce $G_0$. $G_0$ and R produce $G'_0$. Finally, the L-molecules are the precursors of the membrane building blocks M, being transformed in presence of $E_1$ and $E_2$.

**Figure 2 (a)** Space discretization for the lattice model. The grid is formed by squares with a unit surface $dS=dx \cdot dy$. There are tree types of squares or *discrete elements*: *internal elements*, covering the internal space surrounded by the membrane, *external elements* for the area outside the membrane and *membrane discrete elements*. In **(b)** we indicate as $g_{(k,q)-(l,m)}$ the specific flow exchanged between elements (k,q) and (l,m), where the index (l,m) corresponds to different elements around (k,q). When in the neighborhood there are *discrete membrane elements* this values of flow $g_{(k,q)-(l,m)}$ must be corrected, because the flow which arrives to the discrete membrane elements do it along the normal direction defined by the angle $\Phi_{(k,q)}$

**Figure 3.** Time evolution of a model membrane due the effects of the local pressure at each point of the membrane. The simulation starts with a membrane with an arbitrary shape (I) and evolves freely. There are no solutes in the medium, and the shape changes

are due only to the water flow, surface tension, and the membrane elasticity. In this simulation changes in membrane area are due only to its the elasticity, and there no new molecules are added to the membrane. Pictures II, III, and IV show the changing membrane shape at different times. Parameters values from table I but with $R_o$=0.

**Figure 4. C**hanges in membrane area due to the effect of different solute concentrations inside and outside the membrane. The continuous line indicates the evolution as calculated from equation (12). The symbols ( l ) are the numerical results of our simulation method. $\alpha = L_p \cdot P_f / R_f$ where $L_p$ is the hydraulic conductivity, $P_f$ is the final osmotic pressure and $R_f$ is the radius of the membrane at the end.

**Figure 5.** Spatial pressure distribution along the membrane in different times of the simulation. The figures show zones of expansive (positive) and zones of compressive (negative) pressure. This pressure distribution is a consequence of the spatial localization of the metabolic centers and the effects of membrane deformations occurring in these locations. The smaller pictures indicate the membrane shape in each case.

**Figure 6.** Spatiotemporal dynamics of the membrane-metabolism model. The expansion process takes place basically around the metabolic centers $E_1$ and $E_2$ due the osmotic pressure generated by $G_1$. Conversely, in the middle zone the effect of the osmotic pressure associated to R is dominant, and creates a narrowing effect. The plane XY represents the space (as described by our lattice, discrete approximation). The vertical

axis represents the concentration of $G_1$. The maxim is located around the two enzymes (metabolic centers) $E_1$ and $E_2$.

**Figure 7.** Exponential growth of membrane size. The symbols ( l ) indicate the membrane size needed to guarantee a closed, continuous membrane passing through all the characteristic points, which define the membrane shape (see text). The continuous line corresponds to the growth of a membrane from externally available precursors, which are incorporated to the membrane as described by equation (4.2). $T_d$ is the time needed to double the membrane area.

**Figure A.1.** Weight distribution between two membrane *characteristic points* plotted against their relative distance. This profile is similar to the so-called Lennard-Jones potential.

**Figure A.2. (a)** Membrane with a spheroid shape. The interpolation process starts at the *characteristic point* labeled A (this initial point it is arbitrary). The next point with higher weight value, in the clockwise direction, is B, so the interpolation between A and B will be performed. The next interpolation will be between B and C, and so on. (**b)** Here we show a membrane with a very high deformation. In the narrow neck zone there are two possibilities: the interpolation can be performed between B and C, if $W(d_{BC})>W(d_{BC'})$, or between B and C' if $W(d_{BC})<W(d_{BC'})$. In the first case there is only one deformed membrane. In the second case two membranes are in contact.

**Table 1.** Parameter values for the rates of metabolic reactions and membrane deformations used in the simulations displayed in figure 6. The proportionality constant $k_j$ for osmotic pressures in equation (6) is the same for all substances.

**Figure 1**

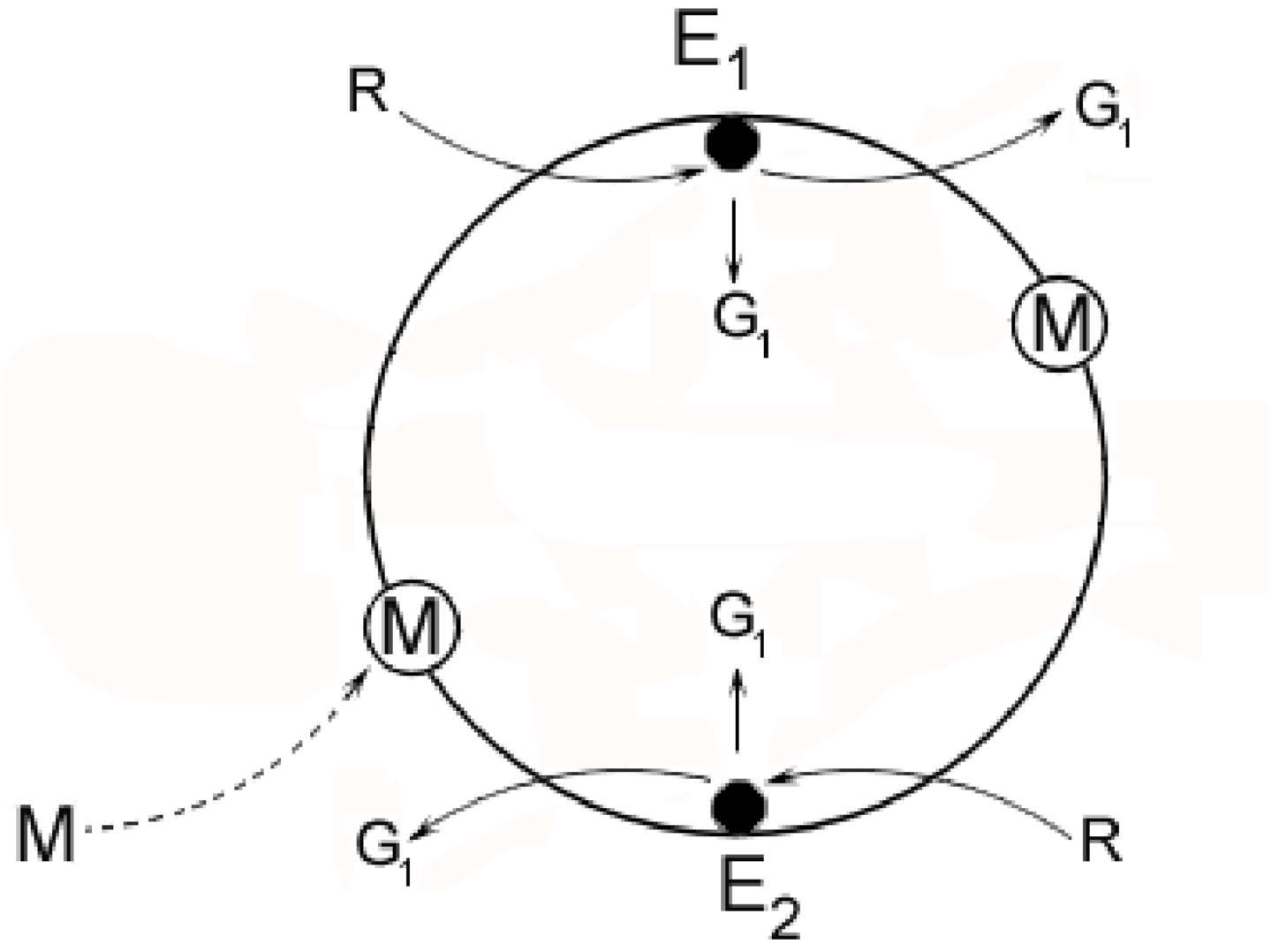

**Figure 2a**

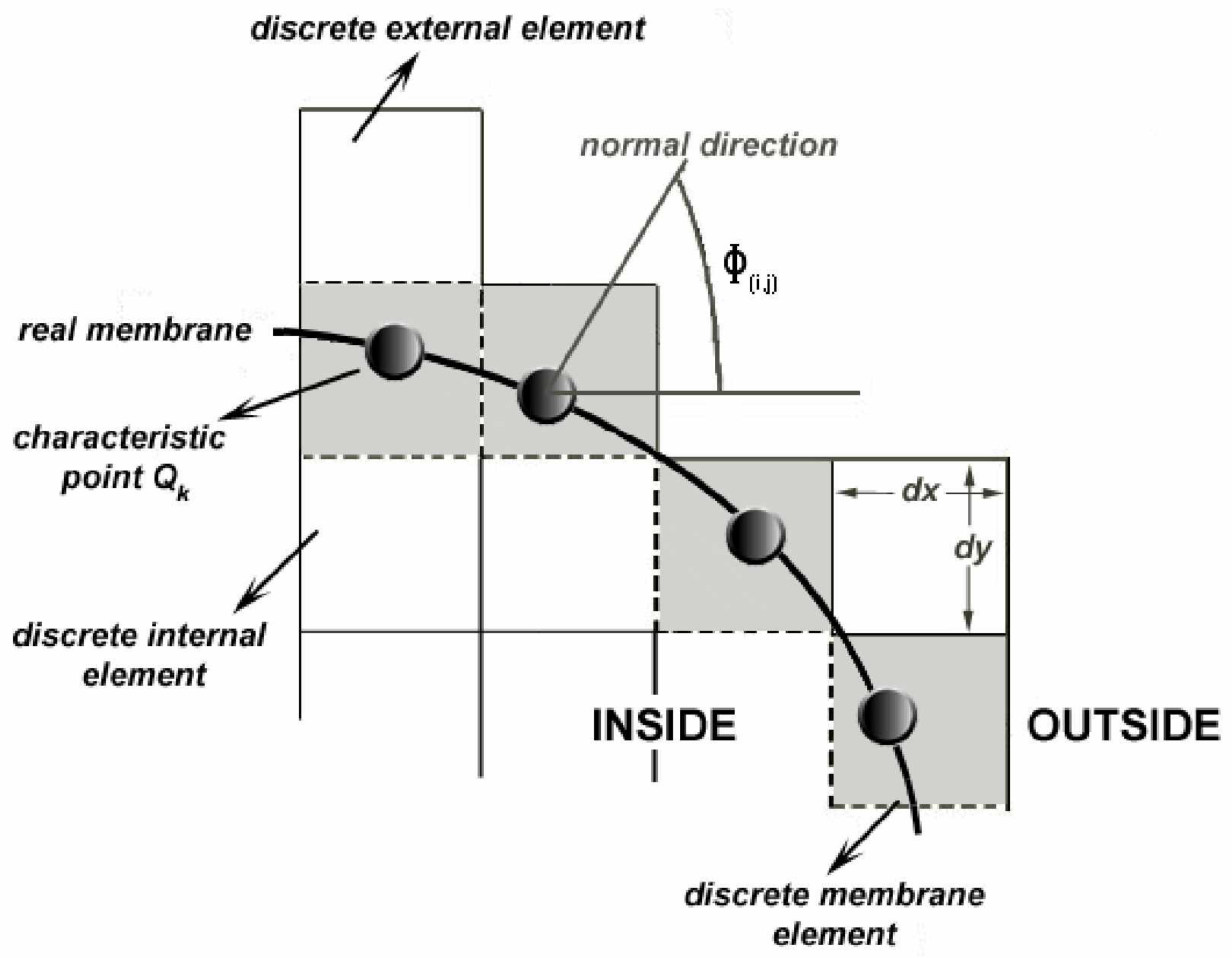

**Figure 2b**

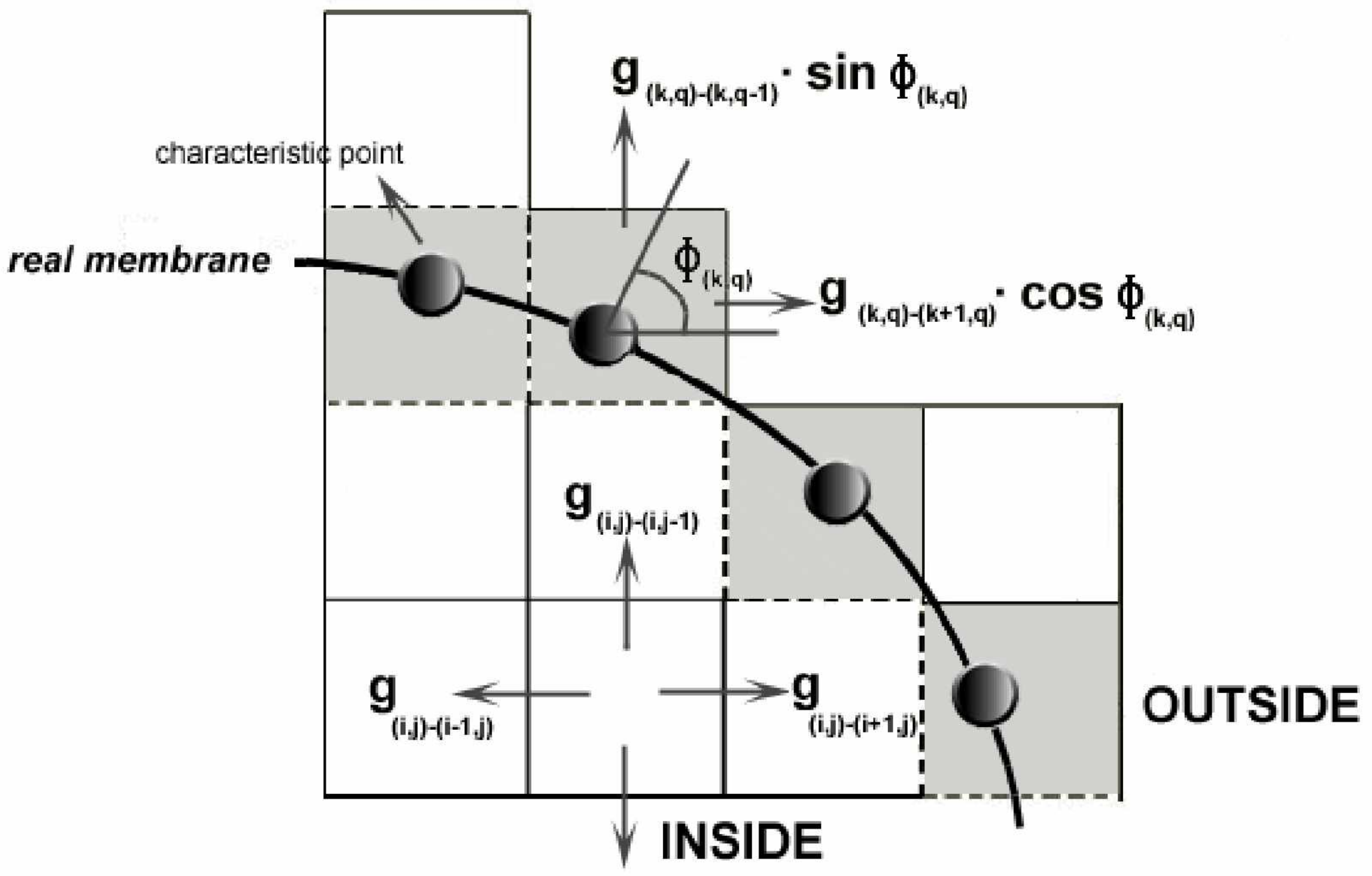

Figure 3

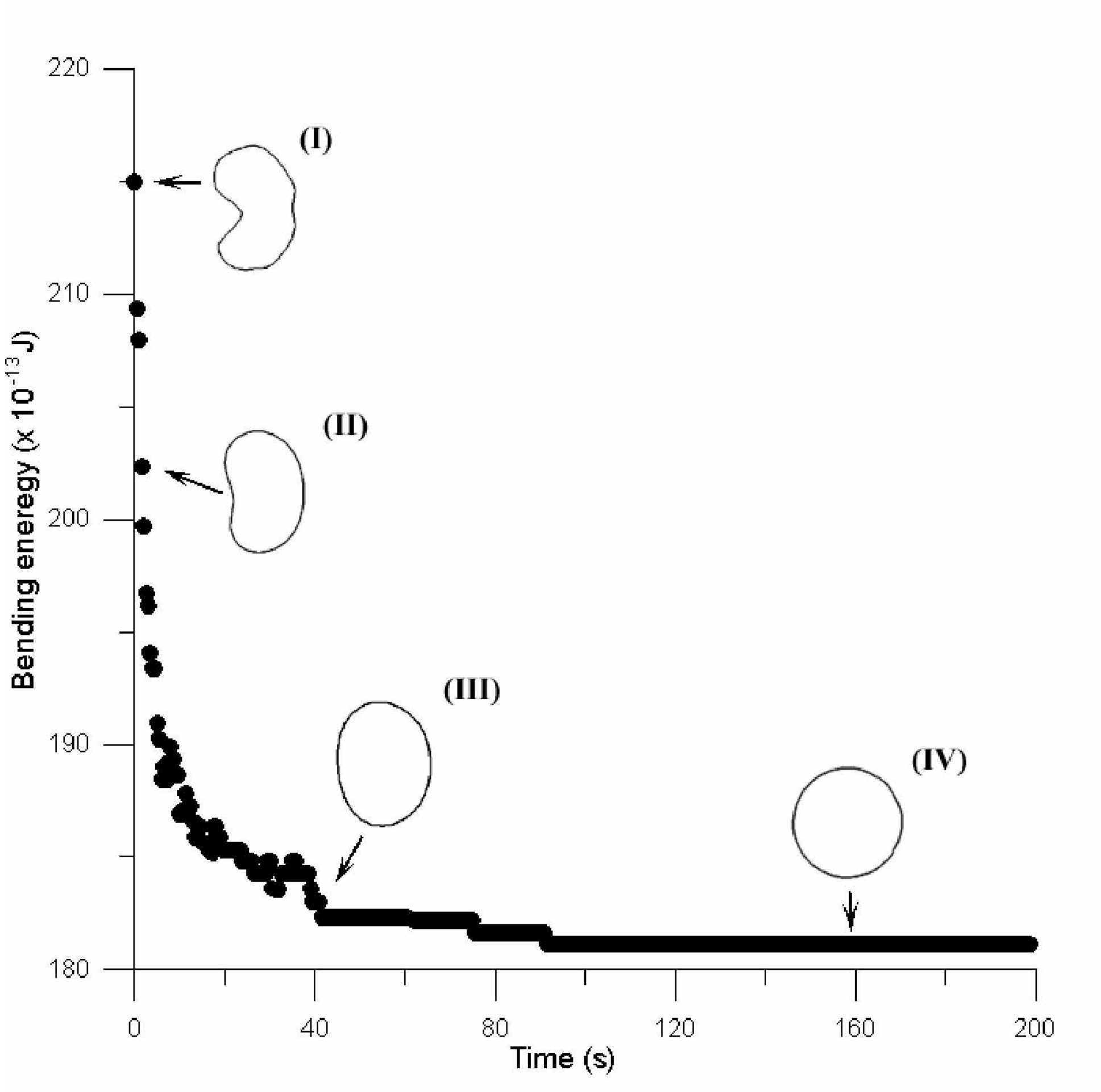

**Figure 4**

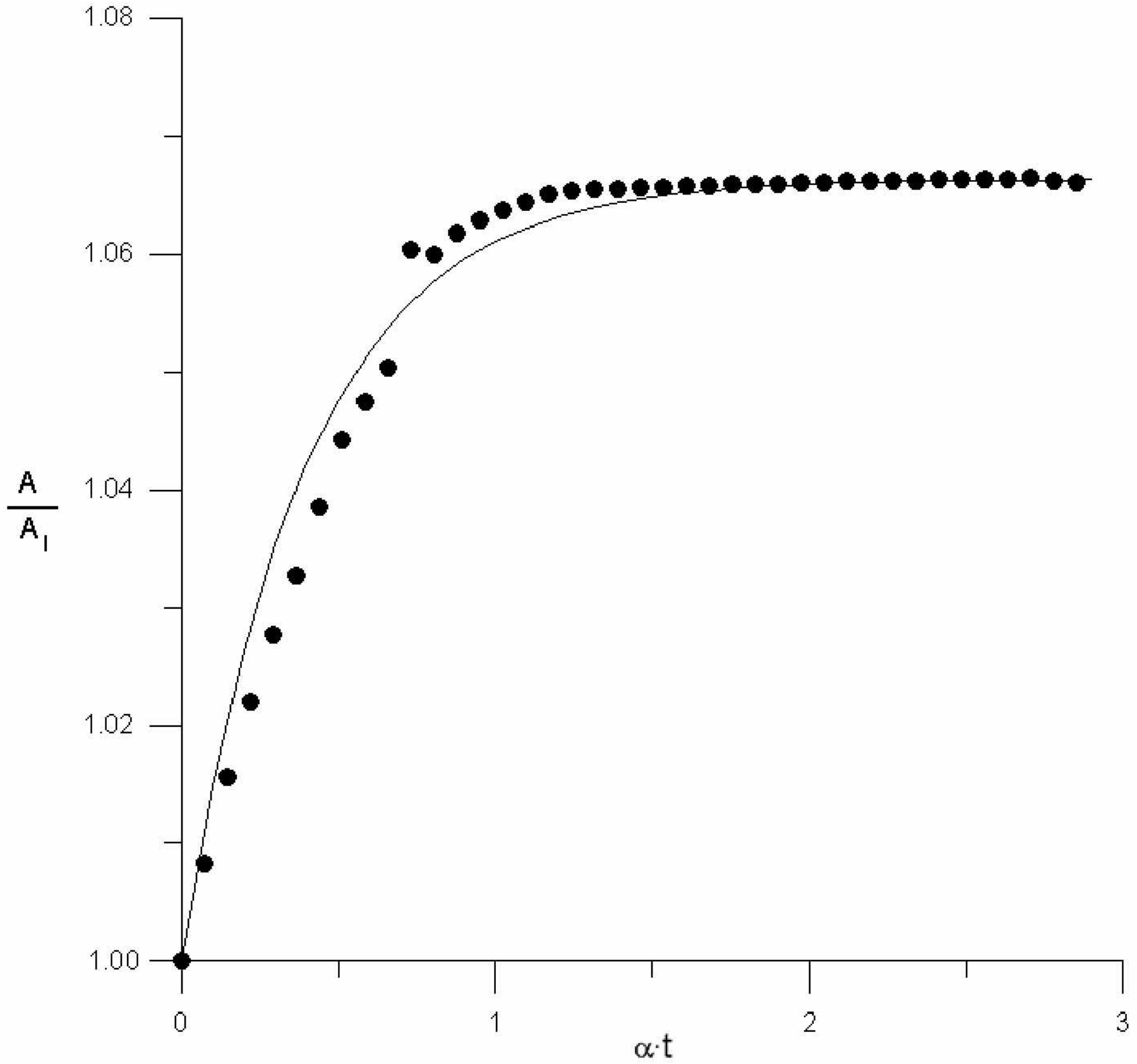

**Figure 5 a-b-c**

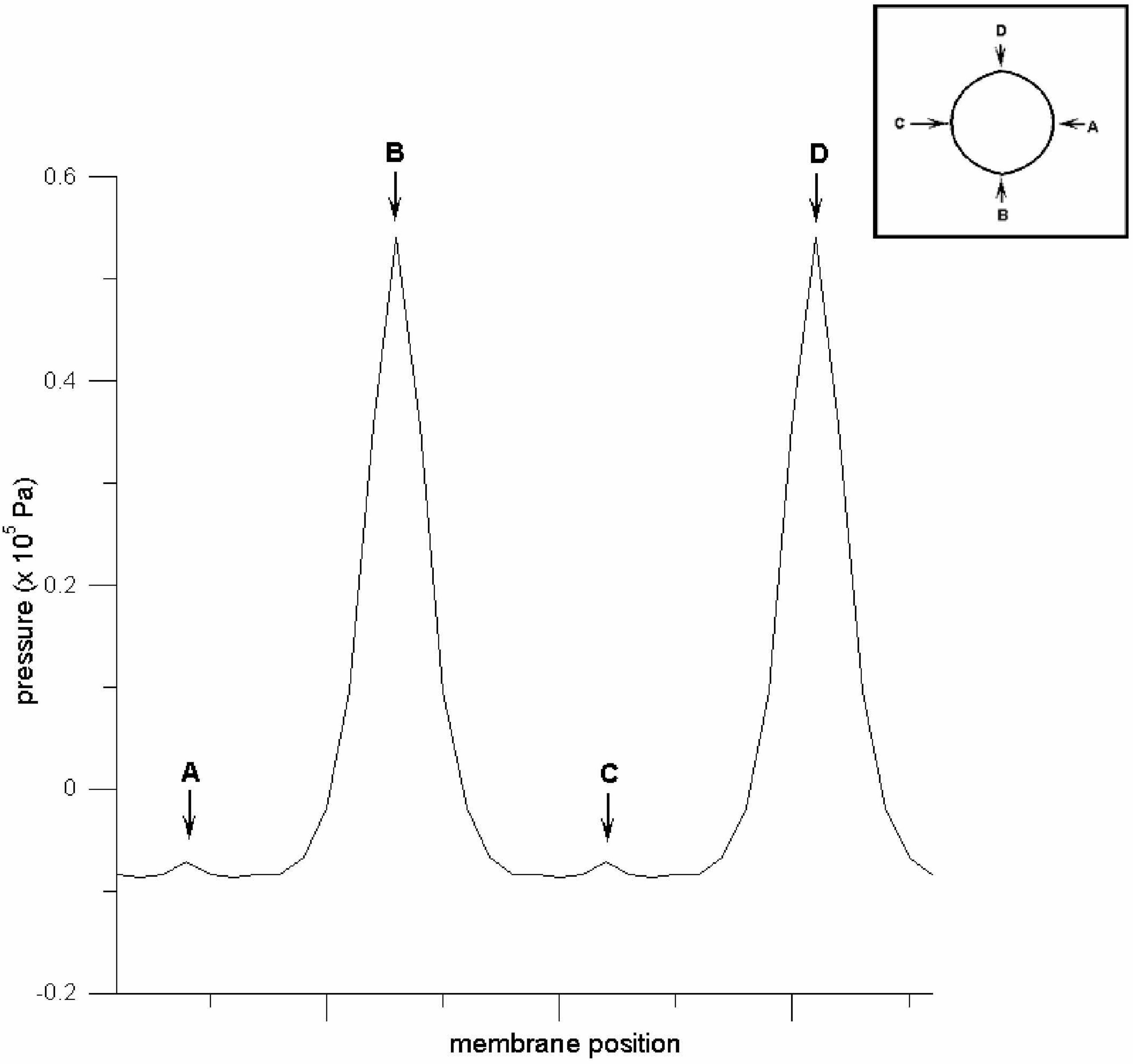

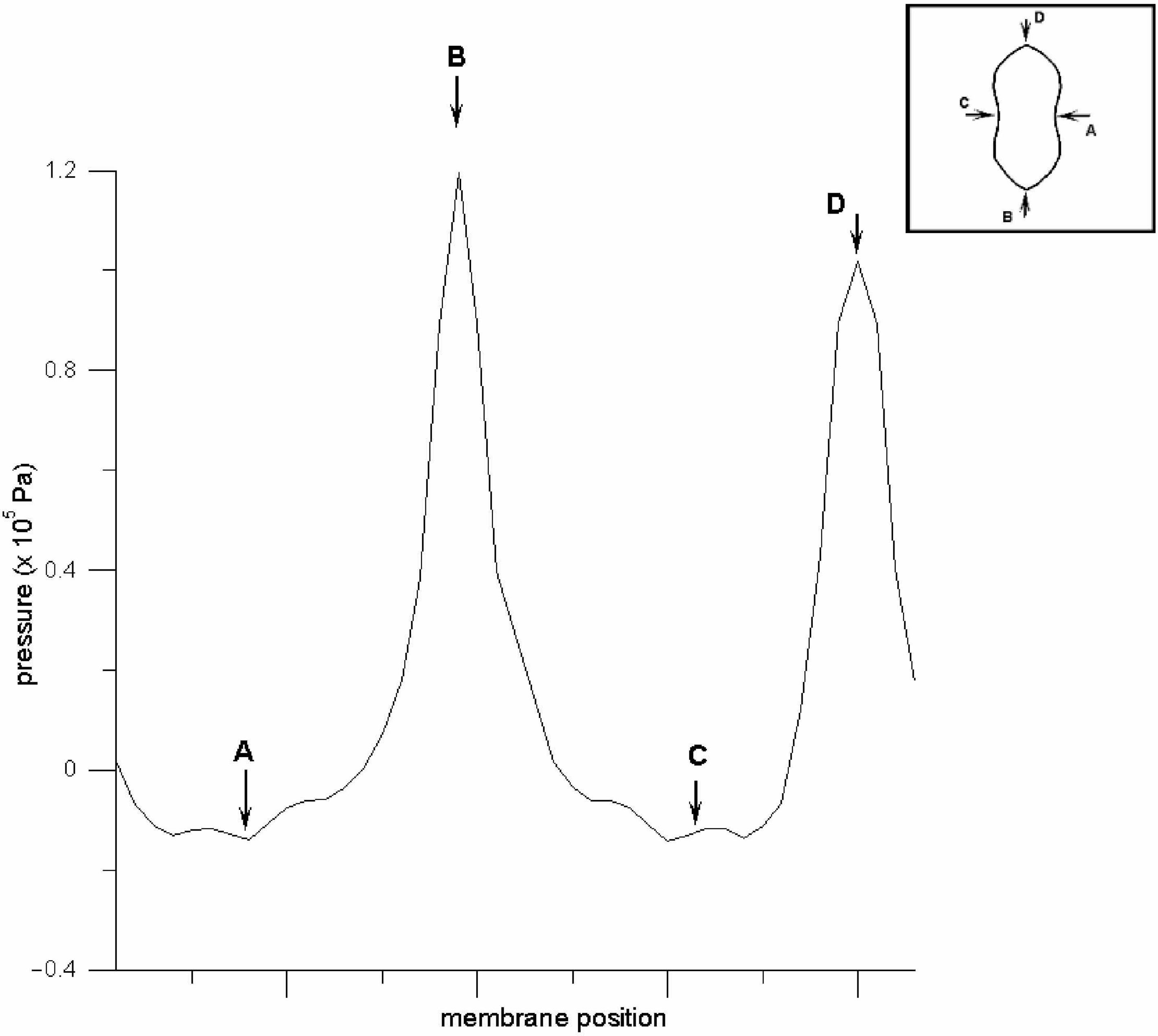

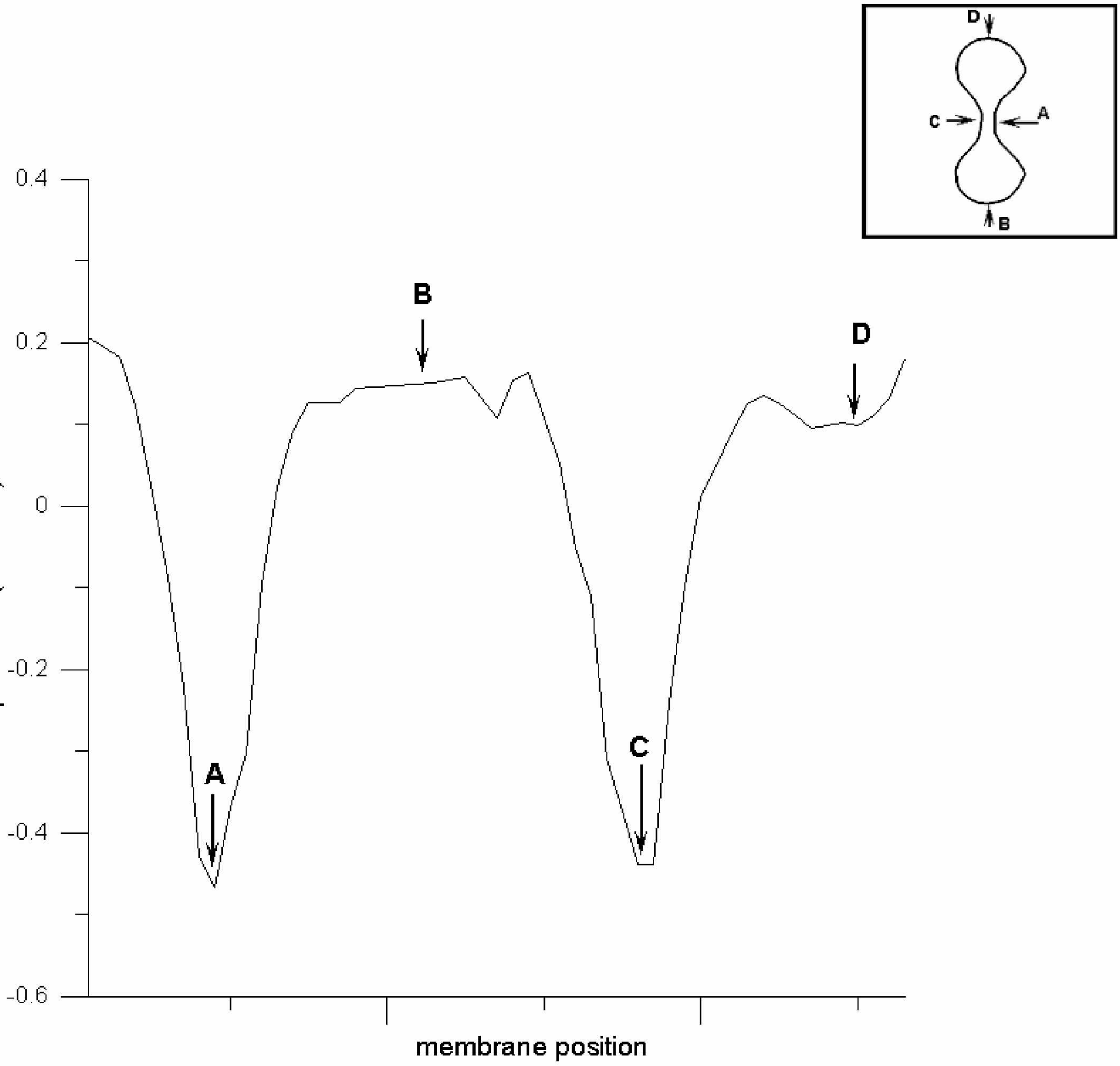

**Figure 6.**

(a) t=15

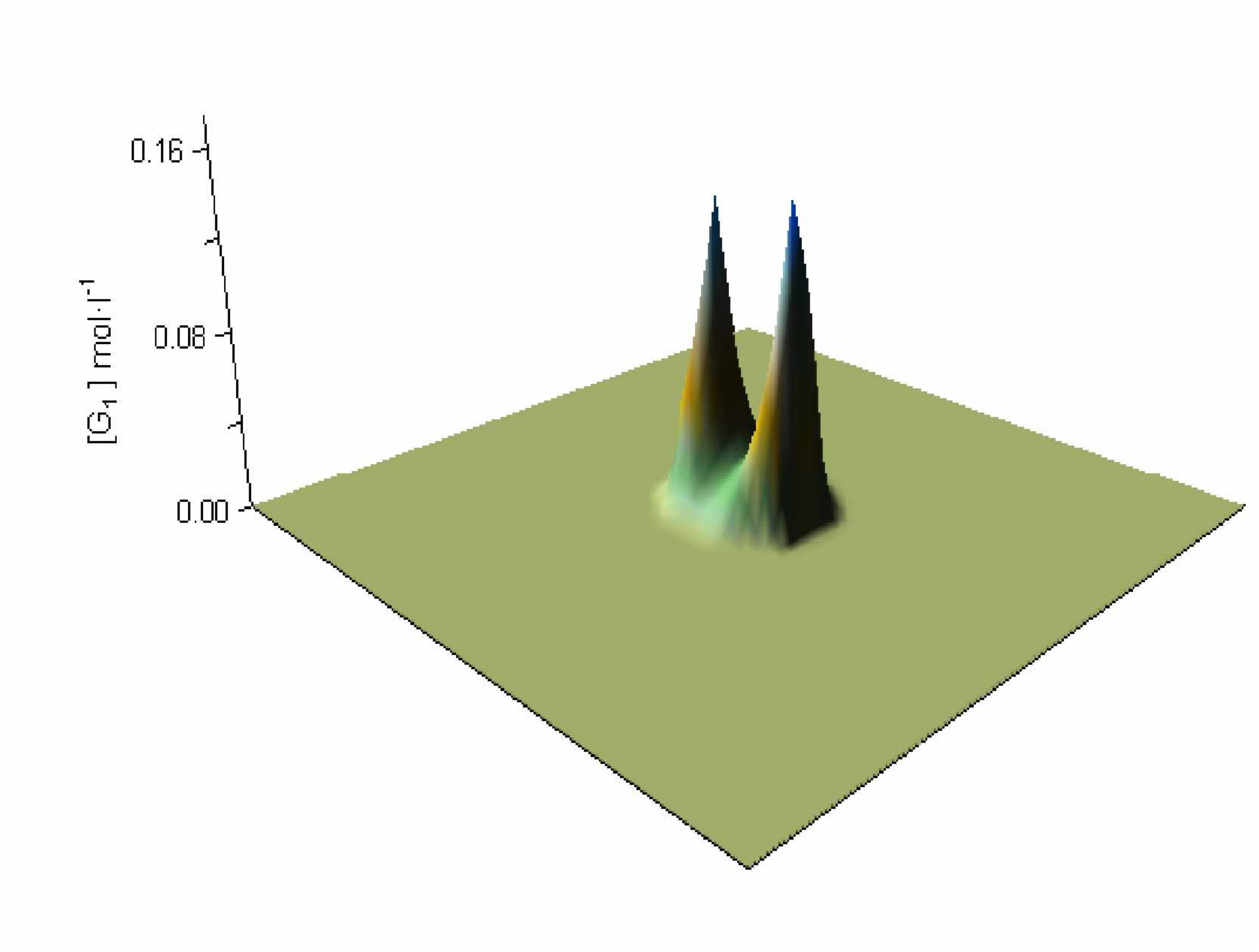

(b) t=90

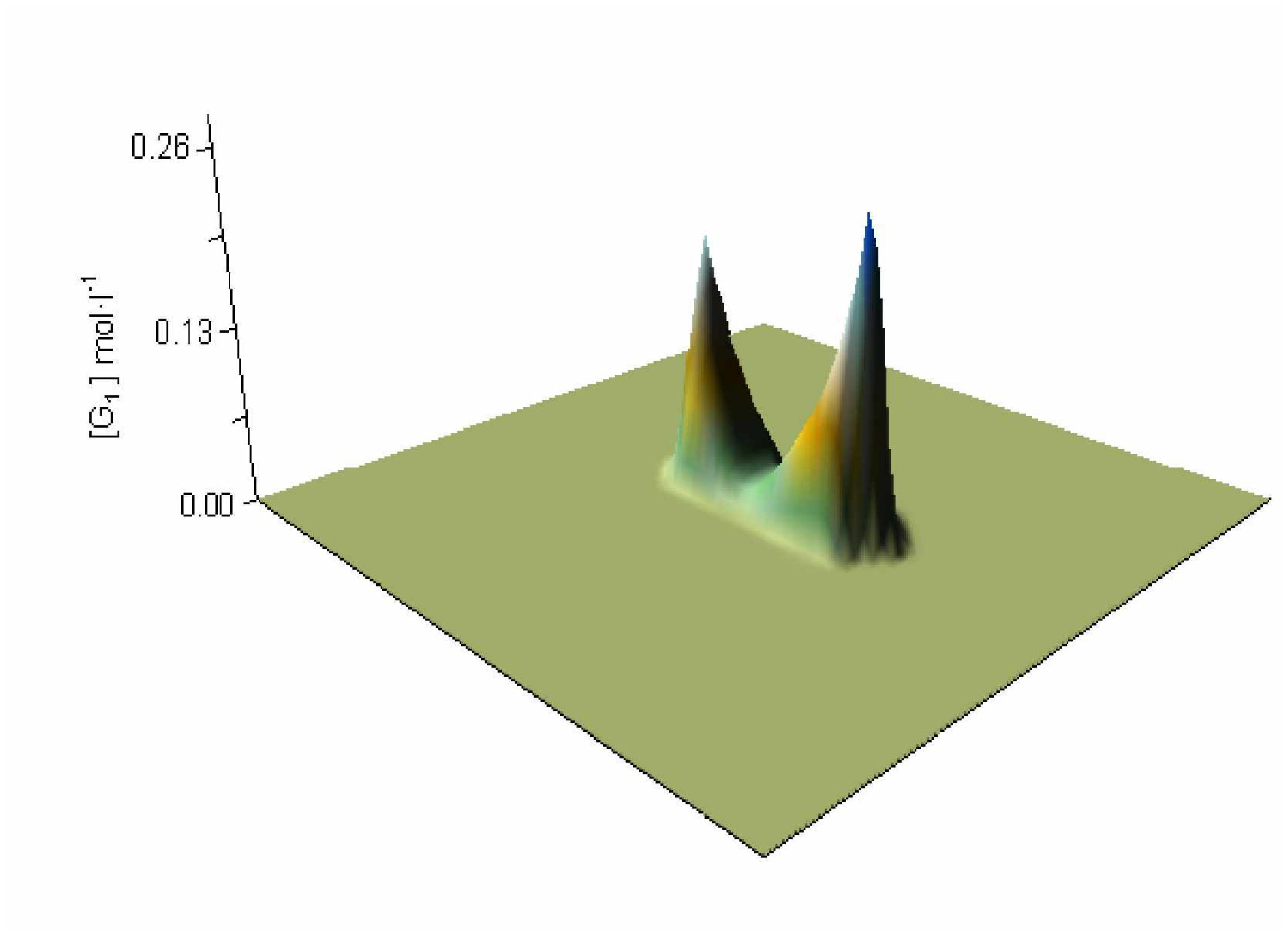

(c) t=170

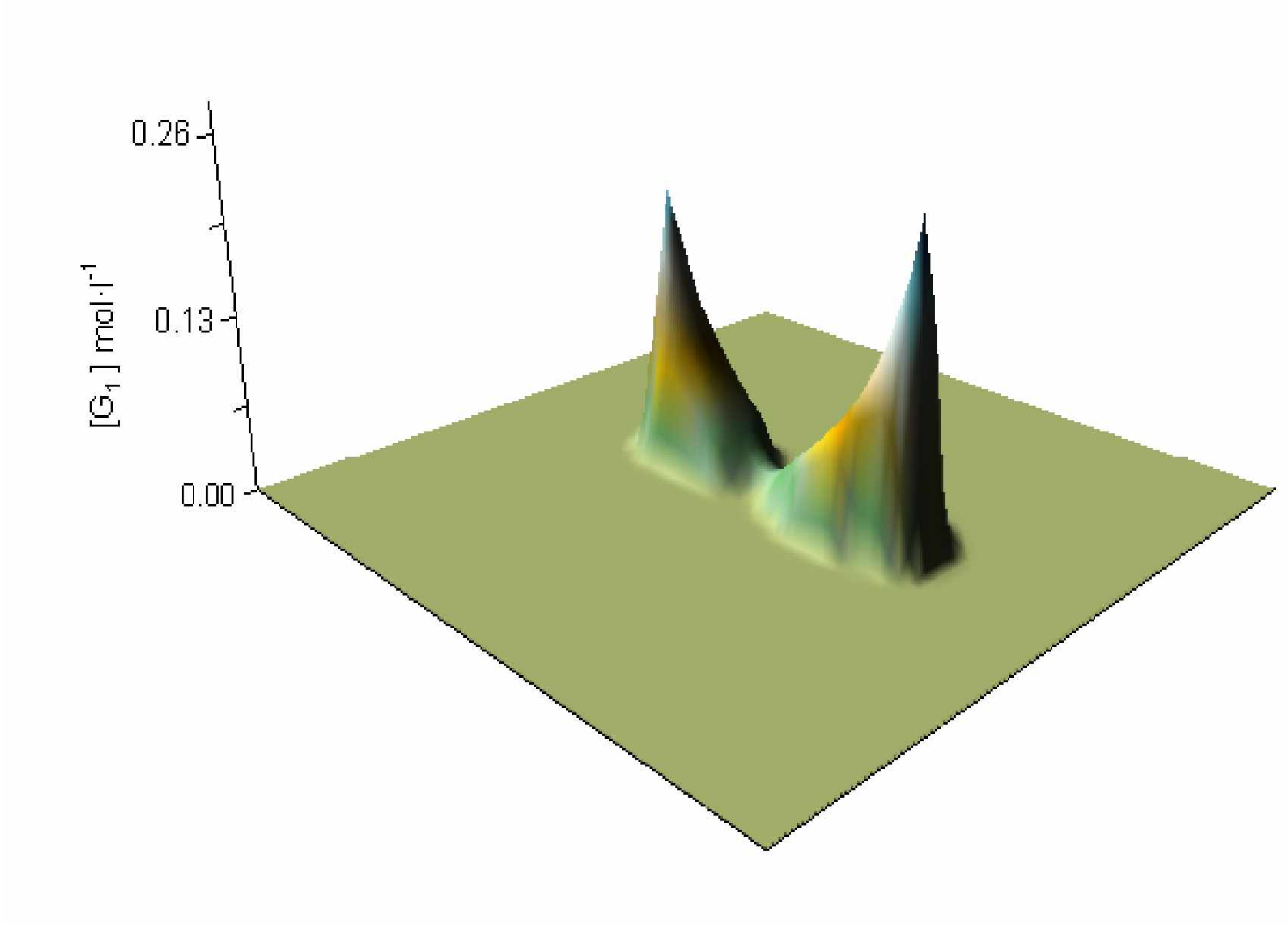

(d) t=220

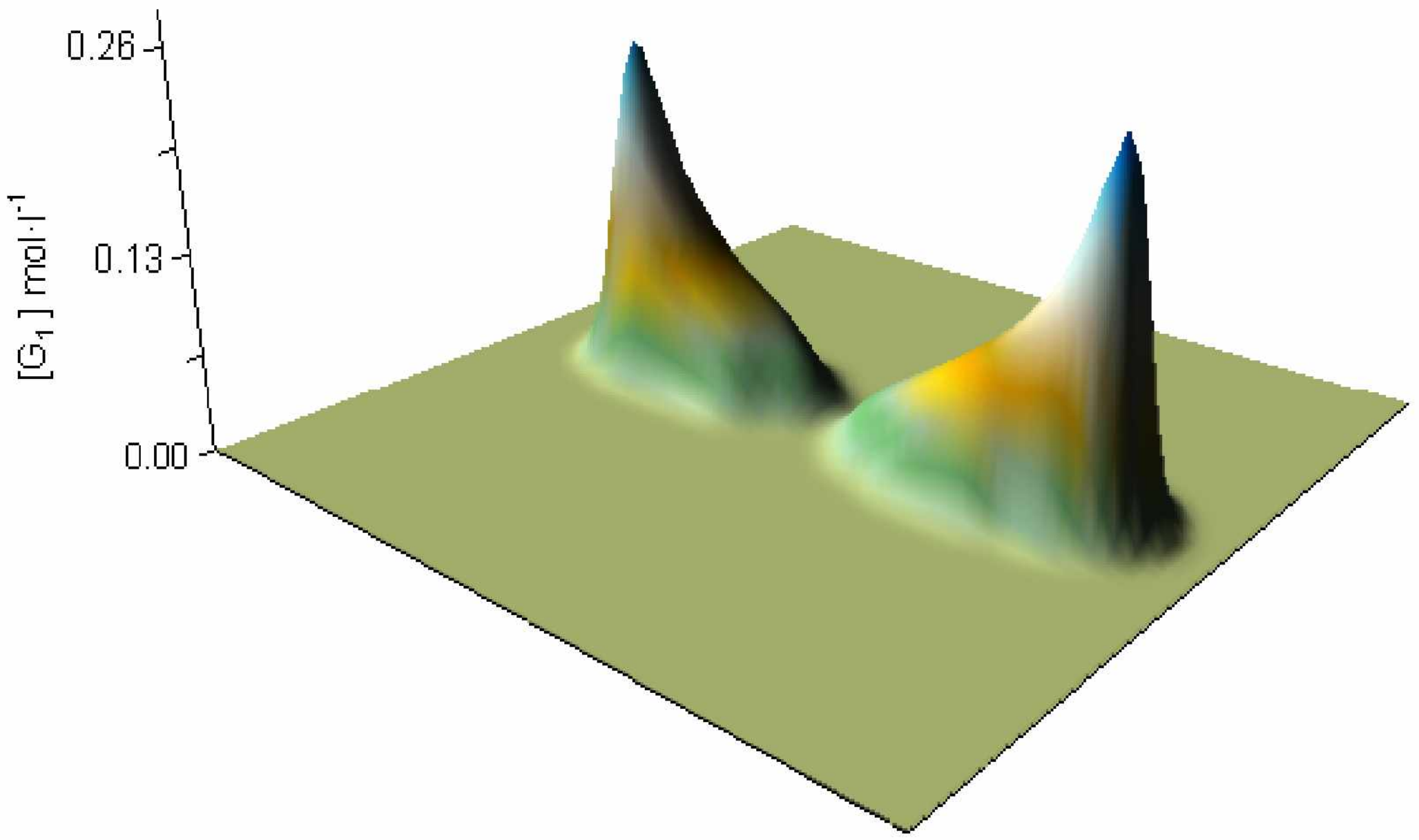
0.26
0.13
0.00
[G1] mol·l-1

**Figure 7.**

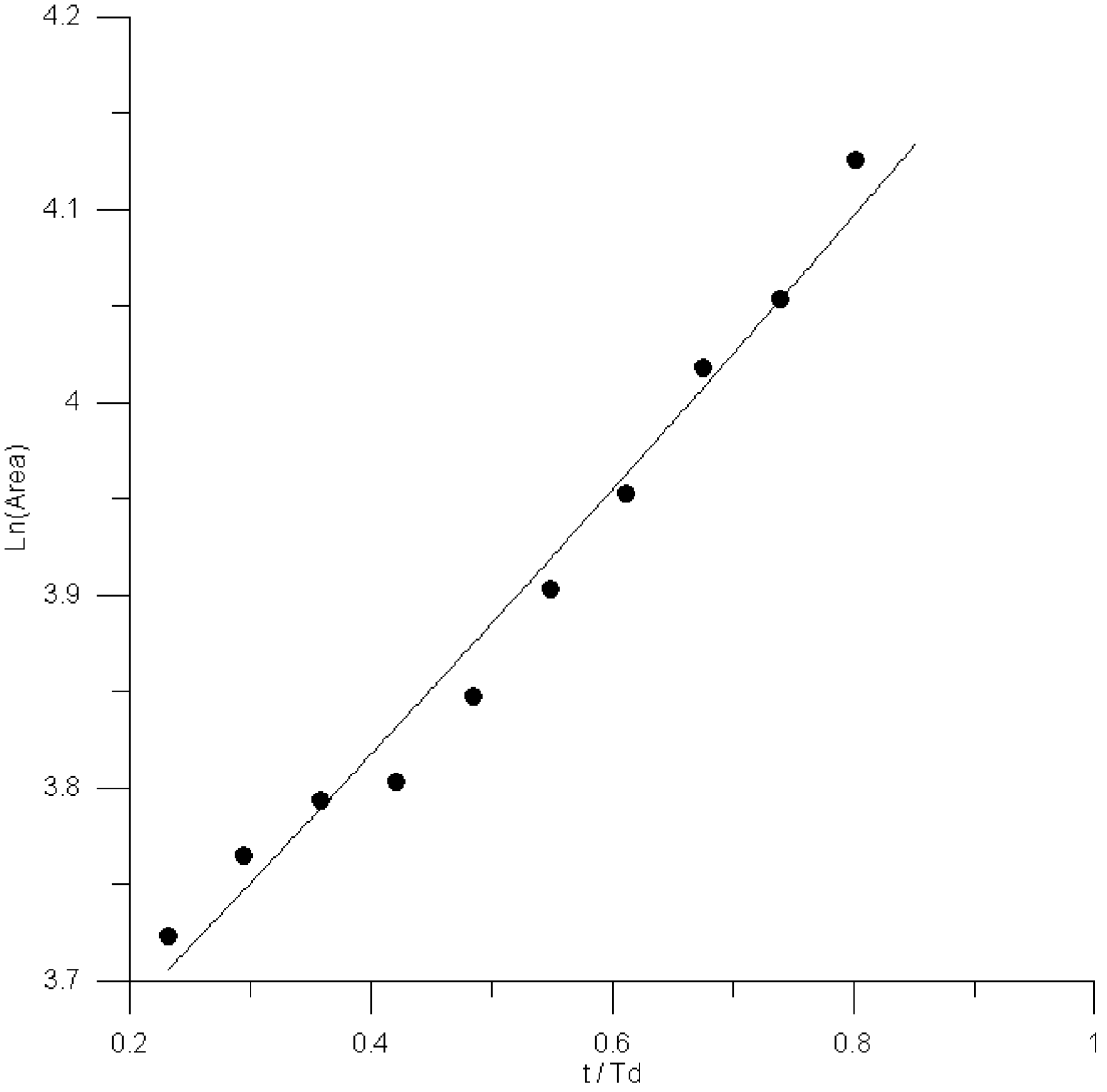

**Figure A.1**

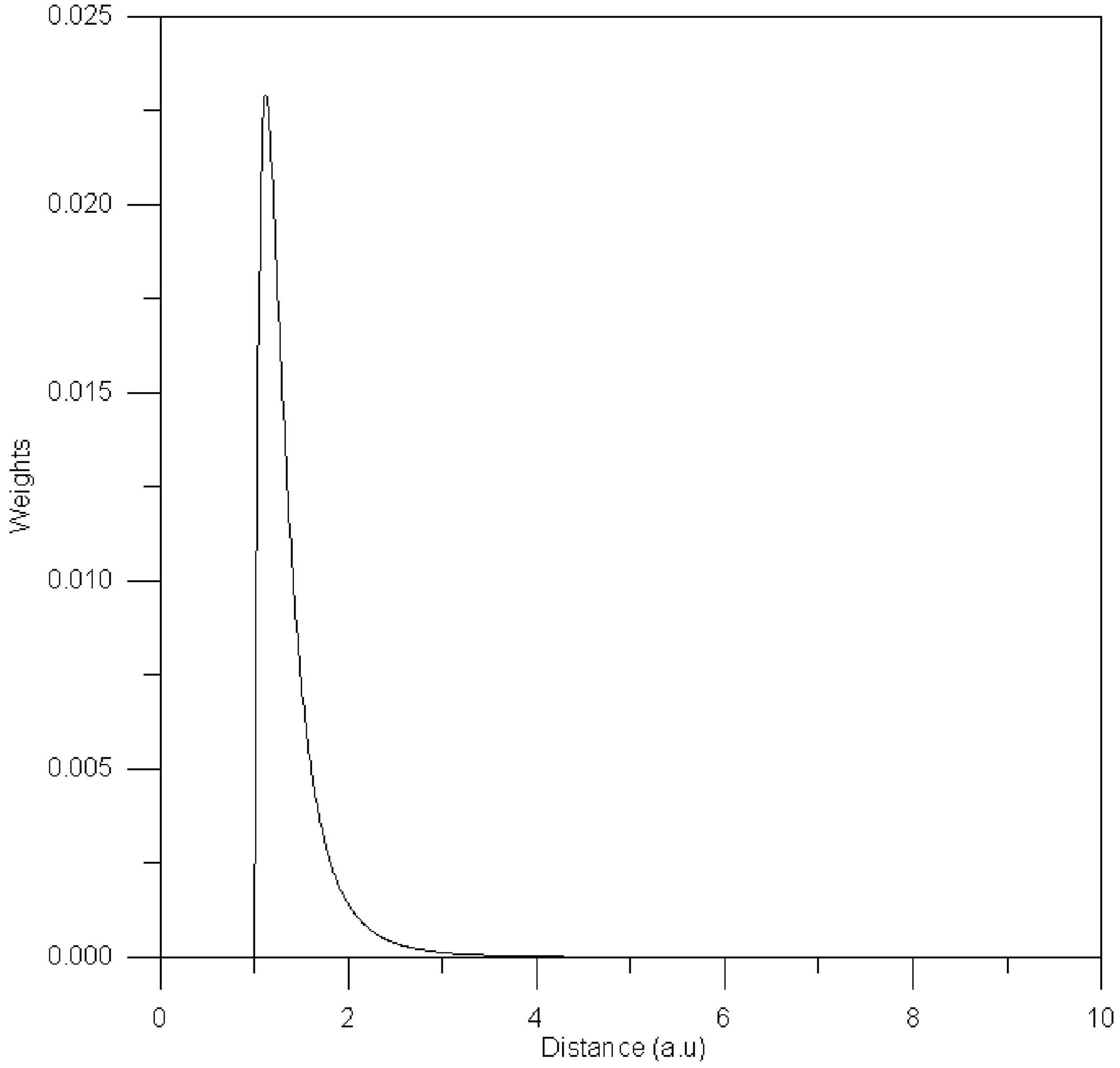

**Figure A.2a-b**

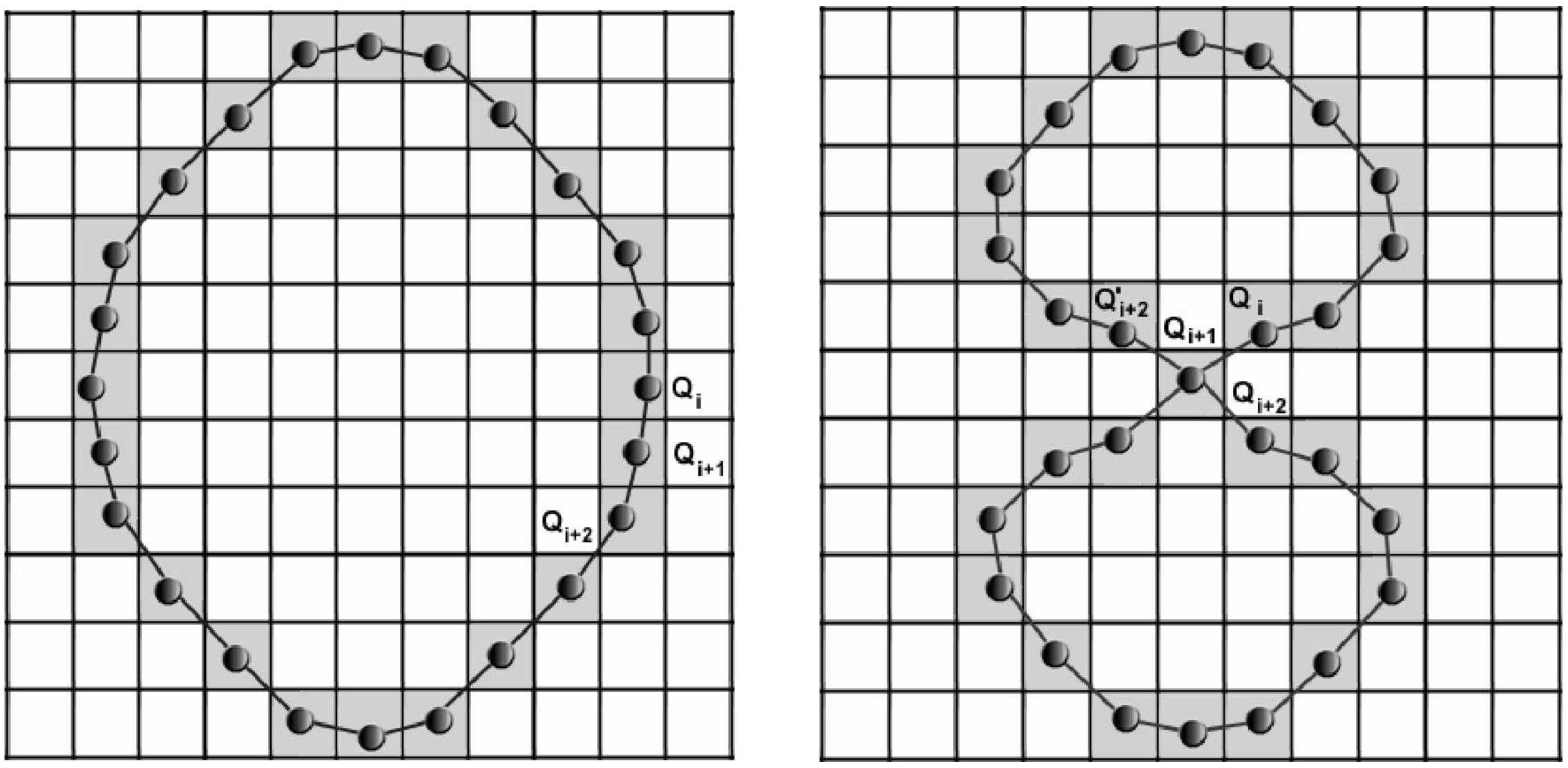

**Table 1.**

| Parameter | Symbol | Value |
|---|---|---|
| Kinetic constant | $k_1$ | $0.7\ M^{-1}\cdot s^{-1}$ |
| “ | $k_2$ | $0.7\ M^{-1}\cdot s^{-1}$ |
| Permeability | $h_R$ | $10^{-8}\ cm\cdot s^{-1}$ |
| “ | $h_{G1}$ | $8\cdot 10^{-8}\ cm\cdot s^{-1}$ |
| Hydraulic conductivity | $L_p$ | $4.1\cdot 10^{-11}\ cm.Pa^{-1}\cdot s^{-1}$ |
| Diffusion coefficient | $D_R$ | $1.6\cdot 10^{-8}\ cm^{2\cdot}s^{-1}$ |
| “ | $D_{G1}$ | $3\cdot 10^{-8}\ cm^{2\cdot}s^{-1}$ |
| Displacement proportionality constant | b | $1.57\cdot 10^{-13}\ cm\cdot Pa^{-1}$ |
| Substance contribution | $R_o$ | $1\ mol\cdot l^{-1}$ |
| Surface Tension Coefficient | $\gamma$ | 2.98 Pa·cm |
| Elastic bending coefficient | *k* | $1.34\cdot 10^{-19}\ Pa\cdot cm^3$ |
| Spontaneous radius of curvature | $r_o$ | 7 µm |
| Temperature | T | 273 K |